\documentclass[aps,pre,preprint,superscriptaddress]{revtex4-1}
\usepackage[latin1]{inputenc}
\usepackage[T1]{fontenc}
\usepackage[english]{babel}
\usepackage{fancyhdr}
\usepackage{graphicx}
\usepackage{float}
\usepackage{amsmath}
\usepackage[math]{blindtext}
\usepackage{natbib}
\usepackage{xcolor}

\begin{document}
        
\title{Recurrence recovery in heterogeneous Fermi--Pasta--Ulam--Tsingou systems}
\author{Zidu Li}
\email{zidu.li@uni-siegen.de}
\affiliation{University of Siegen, Germany}

\author{Mason A. Porter}
\affiliation{University of California, Los Angeles and Santa Fe Institute, USA}

\author{Bhaskar Choubey}
\affiliation{University of Siegen, Germany}


\begin{abstract}

The computational investigation of Fermi, Pasta, Ulam, and Tsingou of arrays of nonlinearly coupled oscillators has led to a wealth of studies in nonlinear dynamics. Most studies {of oscillator arrays} have considered homogeneous oscillators, even though there are inherent heterogeneities between {individual} oscillators in real-world arrays. Well-known FPUT phenomena, such as energy recurrence, can break down in such heterogeneous systems. In this paper, we present an approach {--- the use of structured heterogeneities ---} to recover recurrence in FPUT systems in the presence of oscillator heterogeneities. We examine oscillator variabilities in FPUT systems with cubic nonlinearities, and we demonstrate that centrosymmetry in oscillator arrays may be an important source of recurrence.
\end{abstract}


\maketitle



{\bf {The study of Fermi--Pasta--Ulam--Tsingou (FPUT) arrays has yielded numerous insights into a large variety of nonlinear phenomena, ranging from chaotic behavior to solitary-wave dynamics, breathers, and energy recurrence~\cite{Dauxois2005}. A hallmark of the original FPUT computational experiment is the recurrence of initially excited modes~\cite{Fermi1955,Tuck1972,Pace2019}, but it is difficult to observe such recurrences in laboratory experiments.
A key reason for such difficulties is that the individual elements in an array of oscillators are necessarily nonuniform (through well-defined tolerances even when attempting to manufacture
uniform devices),
which tends to prevent energy recurrence. However, even with the unavoidable heterogeneities from tolerances that are typical of electronic oscillators, we show using numerical simulations that it is possible to arrange the oscillators in an array in a way that recovers a significant amount of energy recurrence. This is especially true for centrosymmetric arrangements, which offer a potential avenue to observe energy recurrence in experimental studies of FPUT arrays. Our results also highlight the importance of accounting for real-world variabilities to understand the behavior of nonlinear systems.}}


\section{Introduction}

Since the original investigation of the one-dimensional (1D) Fermi--Pasta--Ulam--Tsingou (FPUT) system in the 1950s, there have been numerous studies of FPUT arrays~\cite{Fermi1955,Dauxois2005,Flach2005,Miloshevich2009,Midtvedt2014,Lvov2018,Lewis2018,Kuzkin2020}. A 1D FPUT system consists of {an array} of harmonic oscillators that are coupled to each other by nonlinear springs. In the original FPUT numerical experiments~\cite{Fermi1955}, Fermi et al. excited the system in the first (i.e., lowest-frequency) mode of the corresponding linear part of the array, with the expectation that the energy of the first mode would spread to higher modes and eventually lead to equipartition of energy among the modes. 
Instead, however, {they observed} that the energy eventually returned to the first mode after initially spreading to higher modes.
Based on these pioneering numerical computations, it seemed that such energy ``recurrence'' may continue indefinitely. Numerous studies have examined this seemingly paradoxical recurrence~\cite{Zabusky1965,Tuck1972,VanSimaeys2001,Berman2005,Pace2019}, and researchers have even observed very long ``superecurrences'' in  numerical studies of FPUT systems~\cite{Tuck1972,Pace2019}.

Despite the wealth of theoretical and computational studies of FPUT systems, there are very few studies of such systems in laboratory experiments~\cite{Goossens2019,Pierangeli2018}. As discussed in the numerical study of Nelson et al.~\cite{Nelson2018}, one possible reason for this is the inherent variability of objects in real life. It is very difficult to manufacture homogeneous elements for laboratory experiments, so one can expect individual oscillators to have heterogeneities even when one attempts to make them homogeneous. Nevertheless, most studies of FPUT systems have considered homogeneous oscillators arrays, with all oscillators assumed to be identical.

In the real world, it is extremely difficult to build homogeneous elements for a system. For example, in electrical engineering, one constructs oscillators using active and passive components with inherent variability~\cite{James1956,Tripathi2014,Gurnett1957,Tao2018}. Even simple resistors have well-defined tolerances, which one expresses through a strip of popular color codes to designate their values~\cite{James1956,IEC2016}. Similarly, one characterizes transistors in integrated circuits by their variability, which one uses to investigate the best-case and worst-case performance of circuits~\cite{Gurnett1957}. Such variability, which is also often called a ``mismatch'', induces heterogeneities in component values within some bounds. Many prior studies have examined arrays of heterogeneous, nonlinearly coupled oscillators~\cite{Laptyeva2014}. Nelson et al.~\cite{Nelson2018} studied the effects of such oscillator variabilities, which they modeled by incorporating random discrepancies in parameter values, in a 1D FPUT array with a quadratic nonlinearity. They observed that energy recurrence breaks down as variability levels reach typical real-world {values, such as those in electronic systems}. Zulkarnain et al.~\cite{ZULKARNAIN2022} extended these results by conducting a mathematical analysis of the relationship between chaotic behavior and variabilities in FPUT systems.

The presence of typical variabilities in most experimental settings limits the ability to observe recurrences in FPUT systems. Therefore, in the present paper, we explore an approach to recover recurrences in heterogeneous FPUT arrays, despite the presence of variabilities. We demonstrate our approach{, which exploits structured heterogeneities,} using a 1D FPUT system with a cubic nonlinearity (i.e., the FPUT-$\beta$ model).
We analyze the proposed approach to recurrence recovery by examining centrosymmetric FPUT systems with heterogeneous oscillators.

Our paper proceeds as follows. In Section~\ref{array}, we present the equations of motion of a 1D FPUT array and introduce heterogeneities into these equations. In Section \ref{method}, we describe our numerical experiments. In Section~\ref{recovery}, we examine recurrence recovery. In Section~\ref{indiv}, we examine the dynamics of 1D FPUT arrays with individual ``defect'' oscillators. In Section~\ref{centro}, we illustrate that using a centrosymmetric oscillator distribution increases the amount of recurrence. In Section~\ref{conc}, we summarize our results.


\section{Heterogeneous FPUT arrays} \label{array}

\begin{figure}[htbp]
	\centering
	\includegraphics[width=0.9\textwidth]{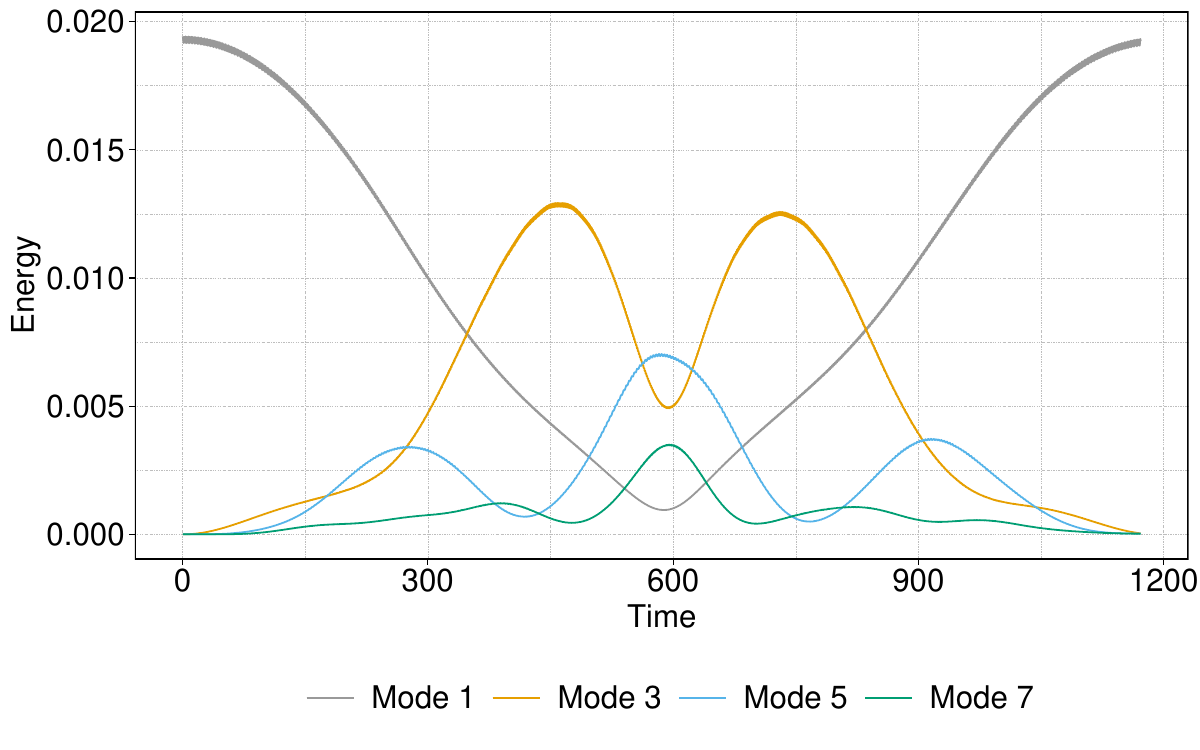}
	\caption{{Energy recurrence in a 1D FPUT array with a cubic nonlinearity}, $N = 128$ oscillators, a nonlinear coupling strength of $\beta = 8$, and no energy in the even modes.
	}
	\label{fig:Idealsituation}
\end{figure}

Consider a 1D array of $N$ harmonic oscillators that are coupled nonlinearly to their nearest neighbors. With a cubic nonlinearity, this yields the FPUT-$\beta$ model
\begin{align} \label{cubic}
	\ddot{x}_1 &= \left(x_{2} - 2x_1\right) + \beta\left[\left(x_{2} - x_1\right)^3 - \left(x_{i}\right)^3 \right] \,, \notag \\
	\ddot{x}_i &= \left(x_{i+1} + x_{i-1} -2x_i\right) + \beta\left[\left(x_{i+1} - x_i\right)^3 - \left(x_{i} - x_{i-1}\right)^3 \right]\,, \quad i \in \{2,3, \ldots, N-1\}\,, \\
	\ddot{x}_N &= \left(x_{N-1} - 2x_N\right) + \beta\left[-\left(x_N\right)^3 - \left(x_{N} - x_{N-1}\right)^3 \right] \,, \notag
\end{align}
where $x_i$ denotes the displacement of the $i$th oscillator from its original position and the parameter $\beta$ describes the coupling strength between neighboring oscillators. In Fig.~\ref{fig:Idealsituation}, we show a simulation that {illustrates the well-known energy recurrence}. In this simulation, in which we excite only the first mode of the system, the initial energy of the first mode spreads to higher odd-order modes, and it then eventually returns to the first mode \cite{Fermi1955}. This cycle then repeats. These simulations use the initial condition $x_i(0) = \sin\left[\pi i/(N+1)\right]$ and $\dot x_i(0) = 0$ (with $i \in \{1, \ldots, N\}$).

The recurrence in Fig.~\ref{fig:Idealsituation} {occurs in} an FPUT array in which all oscillators are identical. However, in the real world, it is extremely difficult (and, in fact, essentially impossible) to build uniform systems. To model a real-world FPUT array, one can introduce tolerances (and hence variabilities) in the individual oscillators of the array. We thus consider the {heterogeneous} FPUT-$\beta$ system
\begin{align} \label{cubictol}
	\ddot{x}_1 &= \left(t_{2}x_{2} - 2t_1 x_1\right) + v_1\beta\left[\left(t_{2}x_{2} - t_1 x_1\right)^3 - \left(t_1 x_{1}t\right)^3\right] \,, \notag \\
	\ddot{x}_i &= \left(t_{i+1}x_{i+1} + t_{i-1}x_{i-1} - 2t_i x_i\right) + v_i\beta\left[\left(t_{i+1}x_{i+1} - t_i x_i\right)^3 - \left(t_i x_{i} - t_{i-1}x_{i-1}\right)^3\right] \,,  \\
		&\qquad \qquad i \in \{2,3, \ldots, N-1\} \,, \notag \\
	\ddot{x}_N &= \left(t_{N-1}x_{N-1} - 2t_N x_N\right) + v_N\beta\left[\left(- t_N x_N\right)^3 - \left(t_N x_{N} - t_{N-1}x_{N-1}\right)^3\right]\,, \nonumber 
\end{align} 
where we now incorporate a tolerance $t_n$ in each oscillator to quantify its mismatch from an ideal device. 


\section{Methodology for our numerical simulations} \label{method}

In our simulations of the heterogeneous FPUT-$\beta$ array~\eqref{cubictol}, we use the initial condition $x_i(0) = \sin\left[\pi i/(N+1)\right]$ and $\dot x_i(0) = 0$, which excites the system in the {first (i.e., lowest-frequency) mode of the corresponding homogeneous linear system}. We use tolerance values from electrical engineering, in which typical components have reported tolerances of $\pm 0.1\%$, $\pm 1\%$, $\pm 5\%$, and $\pm 10\%$~\cite{Horowitz2015}. Manufacturing-induced variabilities generally follow a Gaussian distribution \cite{Nelson2018,James1956,IEC2016}, so we assume that the variabilities in our oscillators follow Gaussian distributions. Again using industry standards from electrical engineering, a $\tau\%$ tolerance signifies that the tolerance $t_i$ has a Gaussian distribution with a mean of $1$ and a standard deviation of $\sigma = (1/3) \times 0.01 \tau$. Therefore, with the $6 \sigma$ bounds {that are} used widely in manufacturing, $99.73\%$ of the values of manufactured devices are in the interval $[1 - 0.01 \tau, 1 + 0.01 \tau]$.

We conduct numerical simulations of the FPUT-$\beta$ system~\eqref{cubictol} by using a 4th-order Runge--Kutta algorithm from the built-in MATLAB code libraries and extending the code of Dauxois et al.~\cite{Dauxois2005}. In a typical cohort of oscillators, the tolerance parameter values are distributed randomly, so a deterministic study is impractical. Therefore, we use Monte Carlo techniques to investigate the effects of oscillator variabilities. We do a large number of simulations of {the system~\eqref{cubictol} with} different parameter values. We draw these parameter values from a Gaussian distribution, which is akin to the typical process that is used in design verification of integrated circuits~\cite{Kim2010}. After simulating {the system}~\eqref{cubictol} 100 times, with different sets of tolerance values $t_i$ with a known {variance}, we plot the energy of various modes in each of these simulations, analogously to {our plot for} the homogeneous FPUT-$\beta$ {system} \eqref{cubic} {in} Fig.~\ref{fig:Idealsituation}. Nelson et al.~\cite{Nelson2018} investigated variability in 1D FPUT {arrays} with quadratic coupling for a variety of array sizes and tolerance values. They observed a breakdown in energy recurrence at typical real-world tolerance values. Therefore, it is unsurprising that we observe a similar recurrence breakdown in 1D FPUT arrays with cubic coupling. 

\begin{figure}[htbp]
	\centering
    	\includegraphics[width=1\textwidth]{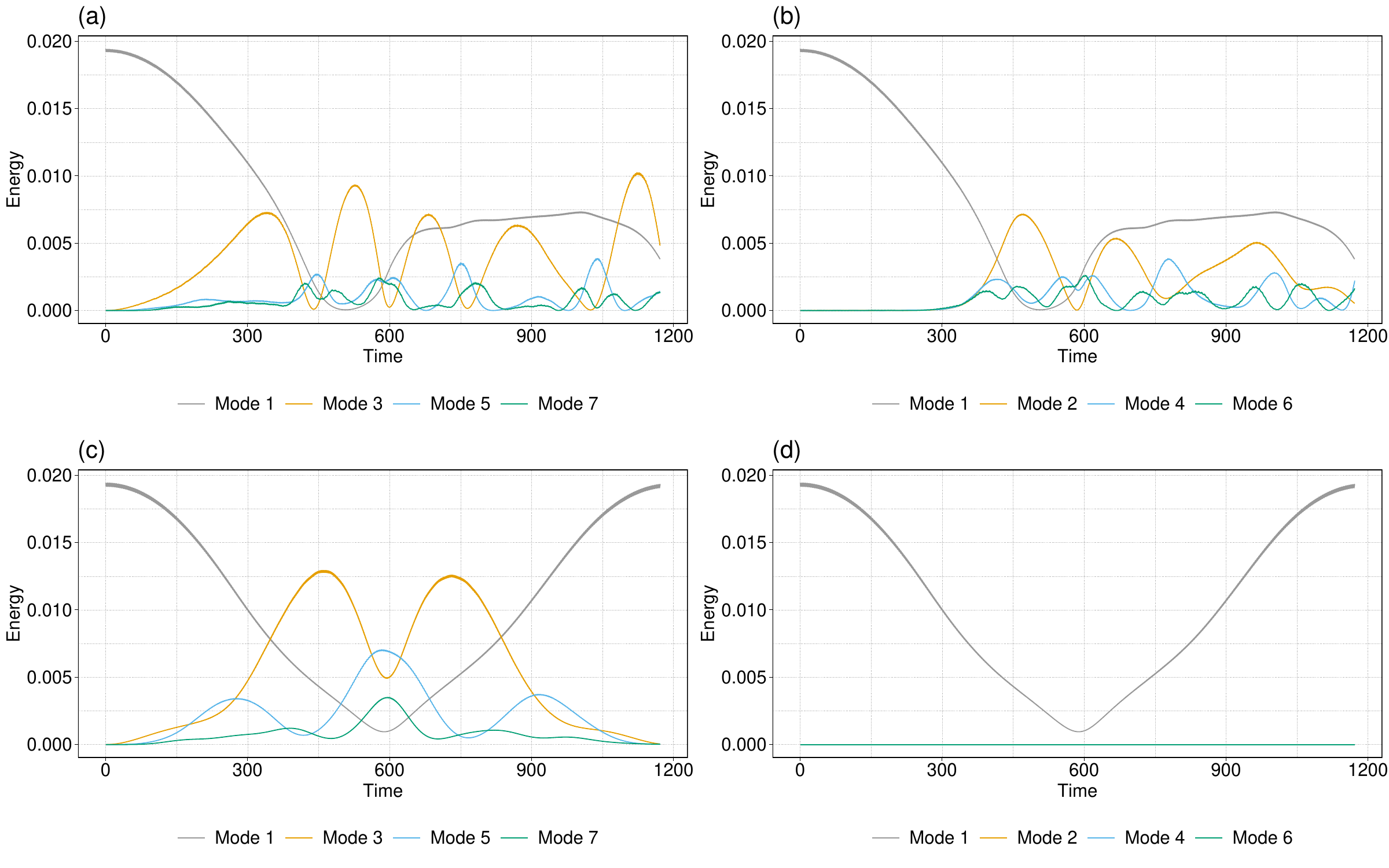}
	\caption{Energy recurrence in a 1D FPUT-$\beta$ array with $N = 128$ oscillators and a nonlinear coupling strength of $\beta = 8$. In {(a, b)}, we show results for a heterogeneous array with a tolerance of $\pm 5\%$; in {(c, d)}, we show results for a homogeneous array.  We show the energy of (a, c) odd modes and (b, d) even modes (with the first mode as a reference). 
	}
	\label{fig:tol101}
\end{figure}

In Fig.~\ref{fig:tol101}(a), we show one typical simulation of the FPUT-$\beta$ array~\eqref{cubictol} with $N = 128$ oscillators, a nonlinear coupling strength of $\beta = 8$, and a tolerance of $\pm 5\%$. In contrast to the homogeneous FPUT-$\beta$ array~\eqref{cubic}, we no longer observe much energy recurrence. This lack of recurrence is characteristic of 85 of the 100 simulations. With tolerances of $\pm 10\%$, only 3 of the 100 simulations of~\eqref{cubictol} still have notable energy recurrence. However, with tolerances of $\pm 0.1\%$ (i.e., very low tolerances), we observe energy recurrence in most of our simulations. 

In a homogeneous 1D FPUT array with cubic coupling, we do not expect even modes to receive energy during recurrence (see Fig.~\ref{fig:tol101}(c,d))~\cite{Fermi1955}. This differs from {a} homogeneous 1D FPUT array with quadratic coupling, where both even and odd modes receive energy during recurrence. {Notably, in} the heterogeneous cubic FPUT array, the even modes also receive energy (see Fig.~\ref{fig:tol101}(b)). This observation is not surprising, but it is still worth illustrating.

\begin{figure}[htbp]
	\centering
	\includegraphics[width=1\textwidth]{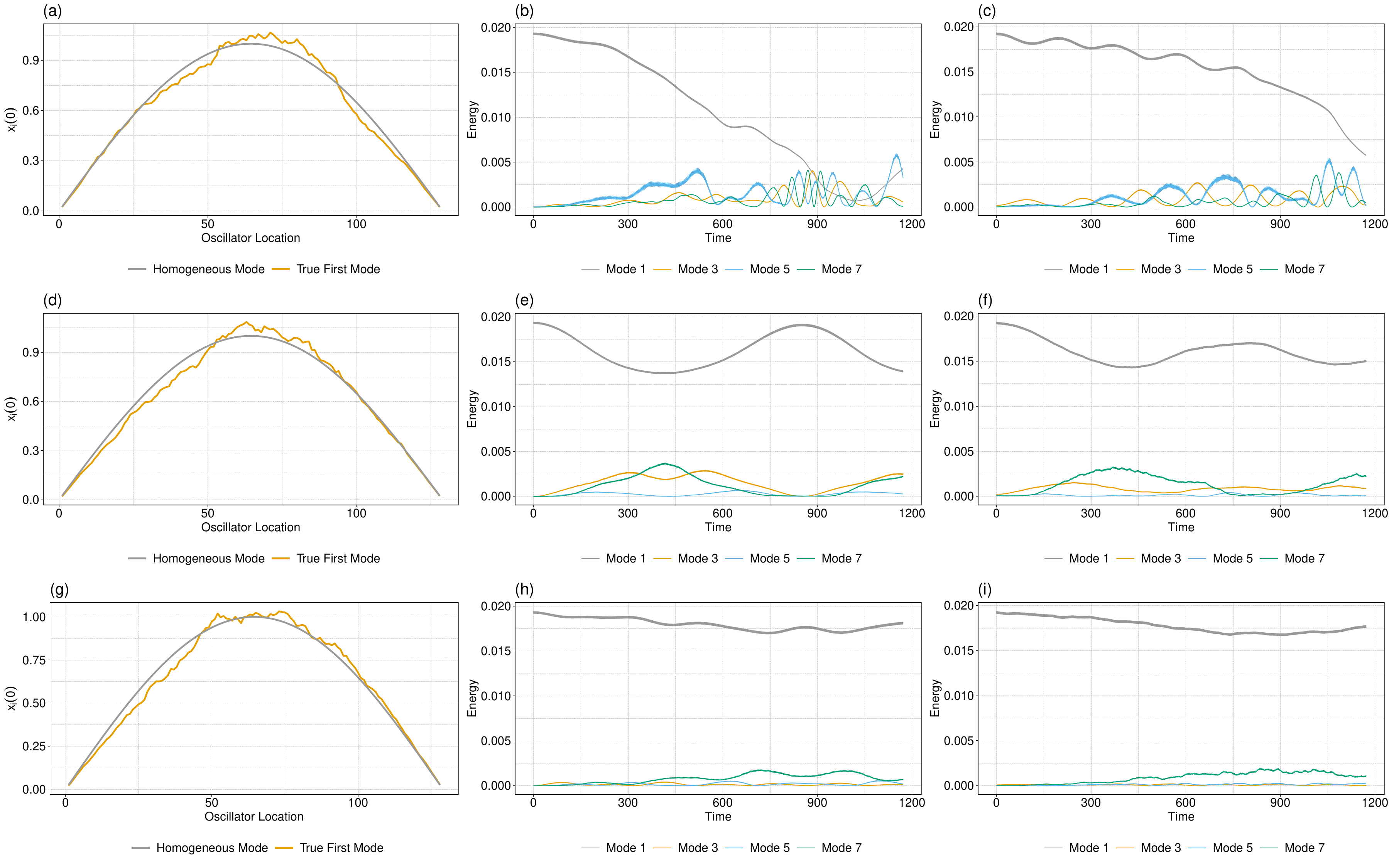}
	\caption{A comparison of energy recurrence in a 1D FPUT-$\beta$ array (with $N = 128$ oscillators, a nonlinear coupling strength of $\beta = 8$, and a tolerance of $\pm 5\%$) for initial conditions that are given by the first true mode of the corresponding linear array and the first mode of the associated homogeneous array.
	In (a, d, g), we show the true first modes (in orange) for three sets of tolerance values and the first mode of the associated homogeneous array (in gray). We denote the oscillator index (i.e., location) by $i$.
	In (b, e, h), we plot the energy behaviors of the system with the first mode of the homogeneous linear array as an initial condition. In (c, f, i), we plot the energy behaviors of the system with the true first modes as the initial condition. Each row corresponds to one set of tolerance values. In the center and right columns, the plotted energies are associated with modes of the homogeneous system.
	}
	\label{fig:nonlinear}
\end{figure}

In both our simulations and in the original study of Fermi et al.~\cite{Fermi1955}, the initial condition is the first mode of the 
linear part of a homogeneous FPUT array. The presence of tolerances in the individual oscillators changes the actual first mode of the corresponding linear array. To investigate if this difference in the first mode causes a breakdown of energy recurrence, we conduct an additional set of numerical computations. We first calculate the true first mode of the corresponding linear system with variabilities. 
We then excite the nonlinear array \eqref{cubictol}, which includes oscillator tolerances, with the corresponding linear system's true first mode. As an example, consider the 1D FPUT-$\beta$ array with $N = 128$ oscillators and a $\pm 5\%$ tolerance. Even when using the true first mode as the initial condition, 92 of the 100 simulations do not exhibit any noticeable difference in its energy profile than what one obtains by exciting the system with the first mode of the corresponding homogenous system. We observe small differences in the exact energy values as the energy changes with time, but we do not observe additional recurrence.
In the other 8 simulations, we do observe a difference in the energy profile; however, 6 of these 8 simulations still do not exhibit energy recurrence. In Fig.~\ref{fig:nonlinear}, we show the results of 3 of these 100 simulations. 

In Figs.~\ref{fig:nonlinear}(a, d, g), we plot both the true first modes and the first mode of the associated homogeneous FPUT array. In Figs.~\ref{fig:nonlinear}(b, c), we plot the energy behavior of a 1D FPUT-$\beta$ array with a $\pm 5\%$ tolerance for these two different initial conditions. We show the energies that are associated with modes of the homogeneous system. In Fig.~\ref{fig:nonlinear}(b), we show the result when the initial condition is the first mode of the homogeneous FPUT array. (See the gray curve in Fig.~\ref{fig:nonlinear}(a).) In Fig.~\ref{fig:nonlinear}(c), we show the result when the initial condition is the true first mode of the heterogeneous FPUT array. (See the orange curve in Fig.~\ref{fig:nonlinear}(a).) Once again, we plot the energies of the modes of the corresponding homogenous system. Similarly, in Figs.~\ref{fig:nonlinear}(e, f) and Figs.~\ref{fig:nonlinear}(h, i), we show the results for the initial conditions in Figs.~\ref{fig:nonlinear}(d, g).

It is not surprising that we do not observe energy recurrence even when we use the true first modes. We expect the nonlinear coupling to play a significant role in recurrence, so initial conditions alone cannot explain the breakdown in recurrence.

\section{Recurrence recovery} \label{recovery}

In our earlier simulations, as well as in those {of Nelson et al.}~\cite{Nelson2018}, there was no predetermined order in the exact locations of the tolerance values in the oscillator array. Indeed, one does not expect any specific ordering of tolerance values in a typical real-world system. We hypothesize that the arbitrariness of the locations of the tolerance values, in tandem with the magnitudes of these values, may be a source of recurrence breakdown. To evaluate this hypothesis, we examine whether or not ordered tolerance values can help restore energy recurrence. This approach of structured heterogeneities was also used by Fraternali et al.~\cite{fraternali2010} in a computational study of {1D arrays} of Hertzian particles.

\begin{figure}[htbp]
	\centering	\includegraphics[width=1\textwidth]{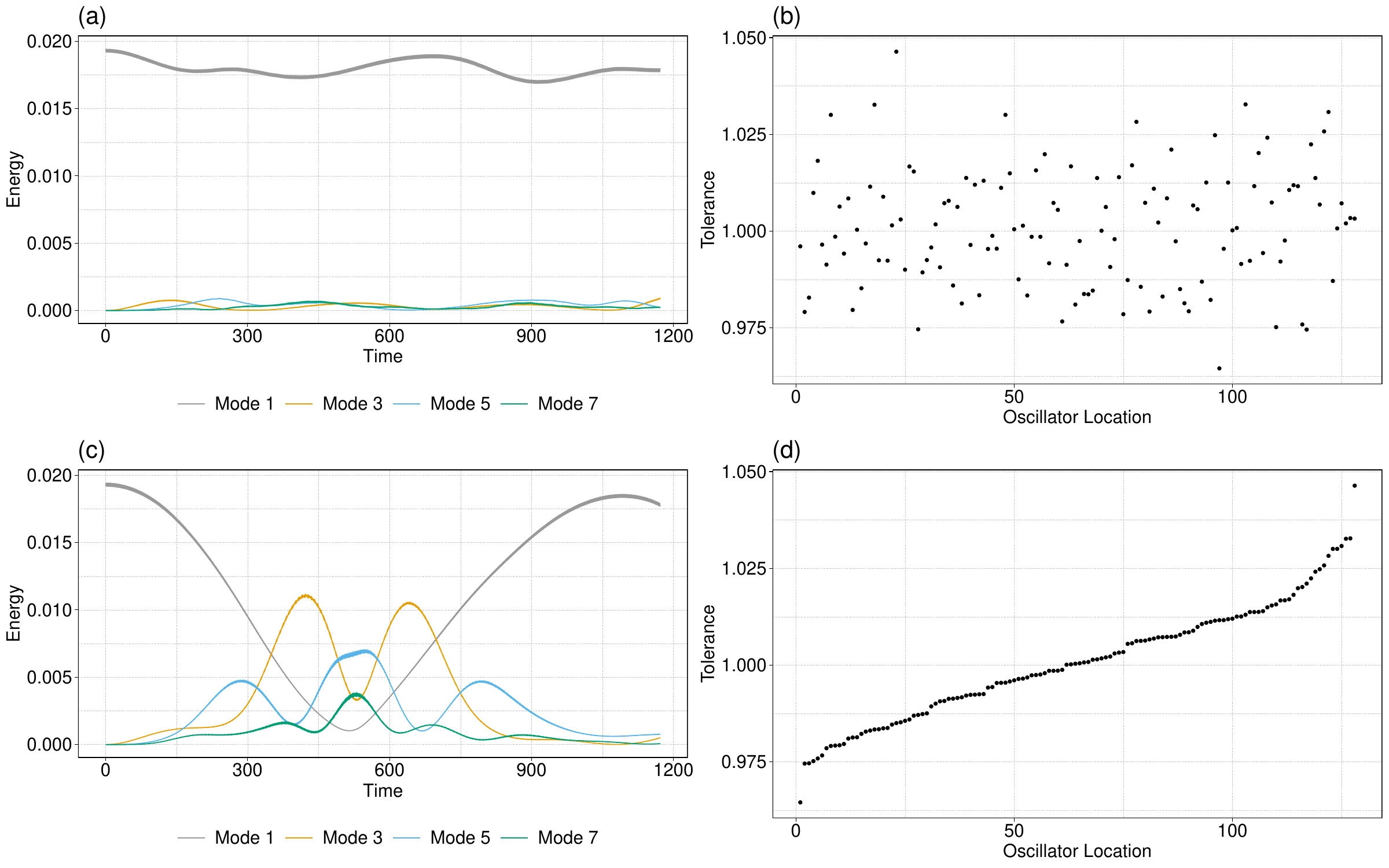}
	\caption{Simulations of 1D {FPUT-$\beta$} arrays with $N = 128$ oscillators, a nonlinear coupling strength of $\beta = 8$, and {a} $\pm 5\%$ tolerance with (a, b) {arbitrarily} located and (c, d) ordered tolerance values.}
	\label{fig:recover}
\end{figure}

We again conduct Monte Carlo simulations of {the system}~\eqref{cubictol}. In Fig.~\ref{fig:recover}(a), we show a typical example {of the energy behavior for an array} with a tolerance of $\pm 5\%$. In Fig.~\ref{fig:recover}(b), we show the locations of {the} individual tolerances values across the 128 oscillators of this FPUT array. To explore our hypothesis {that the order of the tolerance values is important}, we rearrange the oscillator array so that its tolerance values are now in ascending order (see Fig.~\ref{fig:recover}(d)). Simulating this system with the oscillators in this order recovers much of the energy recurrence (see Fig.~\ref{fig:recover}(c)). This result suggests that a key reason for the lack of recurrence may {be} the arbitrary locations of the tolerance values. It also suggests that one should be able to recover much of the energy recurrence of an FPUT array by measuring an appropriate characteristic, such as the tolerance value, of its individual oscillators and then ordering {the oscillators} based on this characteristic. Recall that 85 of the 100 simulations with {arbitrarily placed oscillators with a tolerance of $\pm 5\%$ do} not exhibit notable energy recurrence. {However, by} placing the {oscillators in ascending order of their tolerance values,} we are able to recover recurrence in 23 of these 85 simulations. Consequently, with the ascending order of {tolerance values}, we observe recurrence in 38 of the 100 simulations.

\section{Individual ``defect'' oscillators} \label{indiv}

Our observation of energy-recurrence recovery by rearranging the oscillators of an FPUT array prompts further analysis. In an ordered array, the individual tolerance value of each oscillator depends on its location in the array. Therefore, the terminal oscillators have very different tolerance values {than} those of the center oscillator(s). We expect that the locations of individual tolerance values can affect energy recurrence. We thus numerically simulate 1D FPUT-$\beta$ arrays~\eqref{cubictol} with individual ``defect'' oscillators in an otherwise homogeneous array. The effects of individual and small numbers of isolated defects have also been studied in other nonlinear lattice systems~\cite{cuevas2009,ponson2010,martinez2016}. {If there are sufficiently many defect oscillators, one can obtain nonlinear analogues of Anderson localization~\cite{martinez2016}.}

\begin{figure}
	\centering
	\includegraphics[width=1\textwidth]{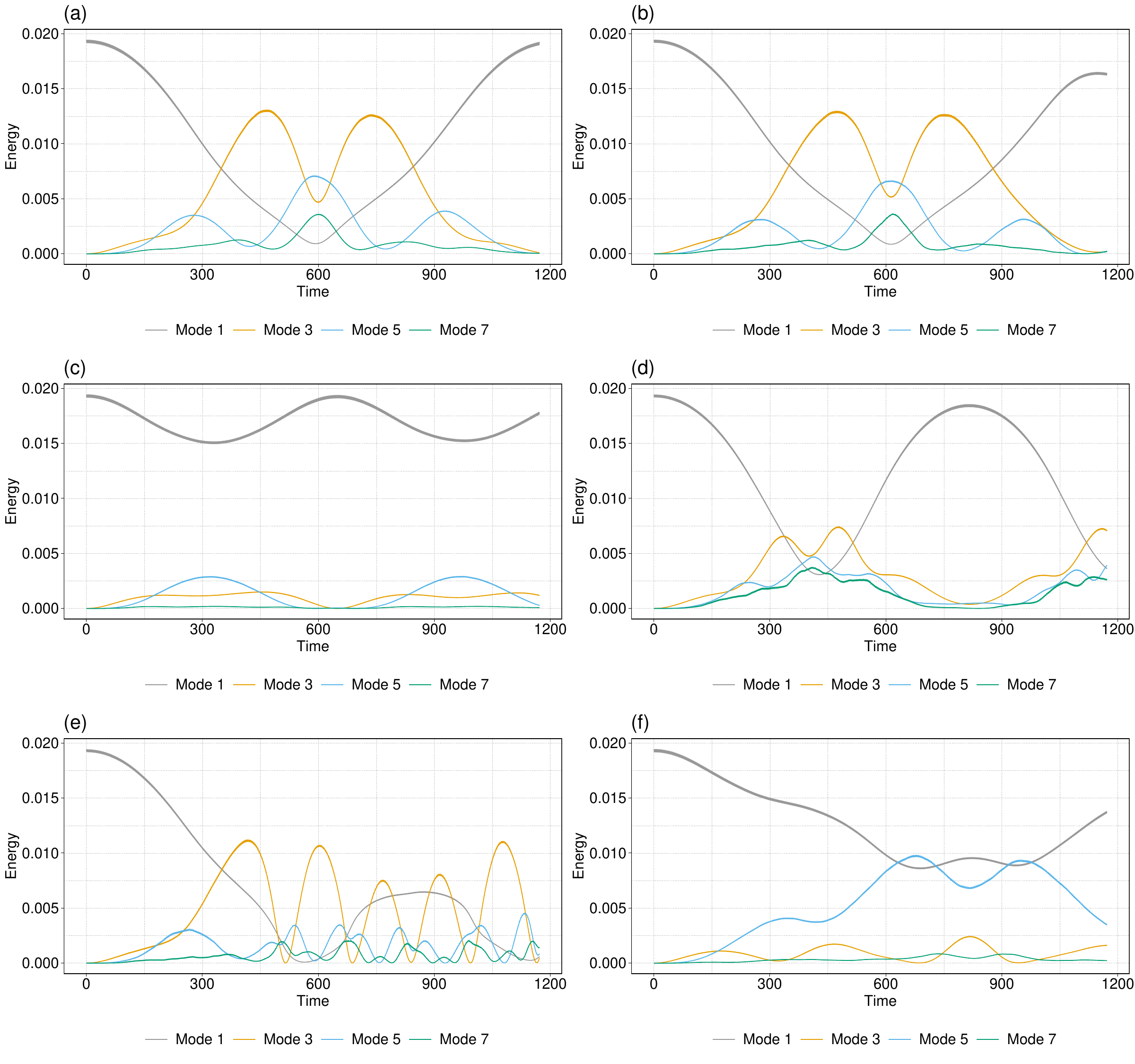}
	\caption{Simulations of a 1D FPUT-$\beta$ array with $N = 128$ oscillators, a nonlinear coupling strength of $\beta = 8$, and $\pm10\%$ tolerance in one of the oscillators.
	We show results for perturbations of (a) oscillator 1, (b) oscillator 32, (c) oscillator 15, (d) oscillator 48, (e) oscillator 61, and (f) oscillator 111.}
	\label{fig:single}
\end{figure}

To examine the effect of {an isolated defect}, we perturb a single oscillator of an FPUT-$\beta$ array. Specifically, we suppose that one oscillator has a nonzero tolerance value and that all other oscillators have tolerance values of $0$. We do numerical computations with a defect oscillator in each of the $N = 128$ oscillators of an array. We again take $\beta = 8$. If the tolerance is sufficiently small, such as at the $\pm 1\%$ level, we observe energy recurrence irrespective of the location of the perturbation. However, if we instead perturb an individual oscillator by giving it a $\pm 10\%$ tolerance, the amount of recurrence breakdown depends on the location of the perturbed oscillator. In Fig.~\ref{fig:single}, we show {the} results of some of our simulations. As we see in Figs.~\ref{fig:single}(a,b), we still observe energy recurrence when we perturb the oscillators at locations 1 and 32. By symmetry, we also observe recurrence when we perturb oscillators 97 or 128 of the array. However, as we can see in Figs.~\ref{fig:single}(c)--(f), we obtain different results when we perturb oscillators in other locations. The recorded loss of recurrence has a different magnitude for different oscillators. This suggests that there may be a location-influenced structure in the system response to oscillator tolerances. This, in turn, influences energy recurrence.

To better understand the energy transfer in our simulations, we want to measure recurrence in some way, as it is impractical to directly visualize the results of all of our simulations. We compute the loss of energy of the first mode in one simulation cycle, which is the time in which we expect to observe recurrence in a homogeneous FPUT array.
When recurrence occurs, the energy in the first mode transfers to higher modes before returning to the first mode. Therefore, as a simple measure of recurrence, we record the difference between the maximum energy and minimum energy of the first mode. In cases without recurrence (see Fig.~\ref{fig:single}(c)), the change in energy of the first mode is small. When there is weak recurrence (see Fig.~\ref{fig:single}(d)), there is a change in the energy of the first mode, but it is smaller than in situations with more significant recurrence. There are also cases with a change in the energy of the first mode but without recurrence (see Fig.~\ref{fig:single}(e)). To account for the diversity of scenarios, we suppose for concreteness that recurrence occurs when at least two thirds of the initial energy of the first mode spreads to higher modes and then the energy of the first mode subsequently returns to at least two thirds of its initial value in one cycle. The choice of the these two thresholds of two thirds is arbitrary; one can instead employ any other threshold values of energy transfer and recovery. In Fig.~\ref{fig:energychange}, we show the profile of this energy difference for FPUT-$\beta$ arrays with {a single defect oscillator.}

We see that perturbing the oscillators at locations 1, 32, 97, and 128 of the array does not lead to a breakdown in recurrence. This is interesting but unsurprising, as these locations are the nodes of the second and fourth modes of a corresponding linearly coupled array. Ideally, the even modes of a homogeneous 1D FPUT-$\beta$ array should not receive energy during recurrence. Therefore, we expect that perturbing the oscillators at the nodes of the modes should not have a major effect on energy recurrence. Our observations also suggest that perturbing the system at some locations may significantly increase the energy transfer to higher modes. This may help explain the breakdown of energy recurrence.


\section{Centrosymmetric distributions of oscillators} \label{centro}

\begin{figure}[htbp]
	\centering
	\includegraphics[width=0.8\textwidth]{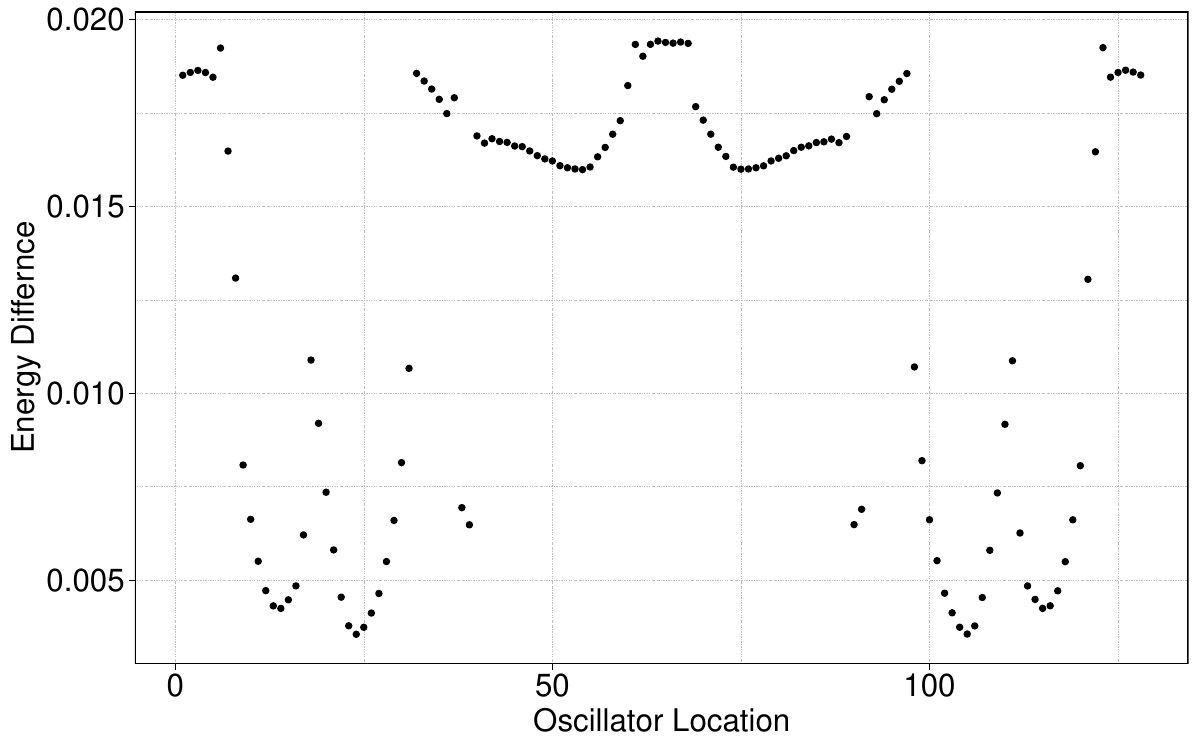}
	\caption{Energy transfer from the first mode for a $\pm 10\%$ tolerance perturbation at different locations of a 128-oscillator 1D FPUT-$\beta$ array with a nonlinear coupling strength of $\beta = 8$.}
	\label{fig:energychange}
\end{figure}

\begin{figure}[htbp]
	\centering
	\includegraphics[width=0.8\textwidth]{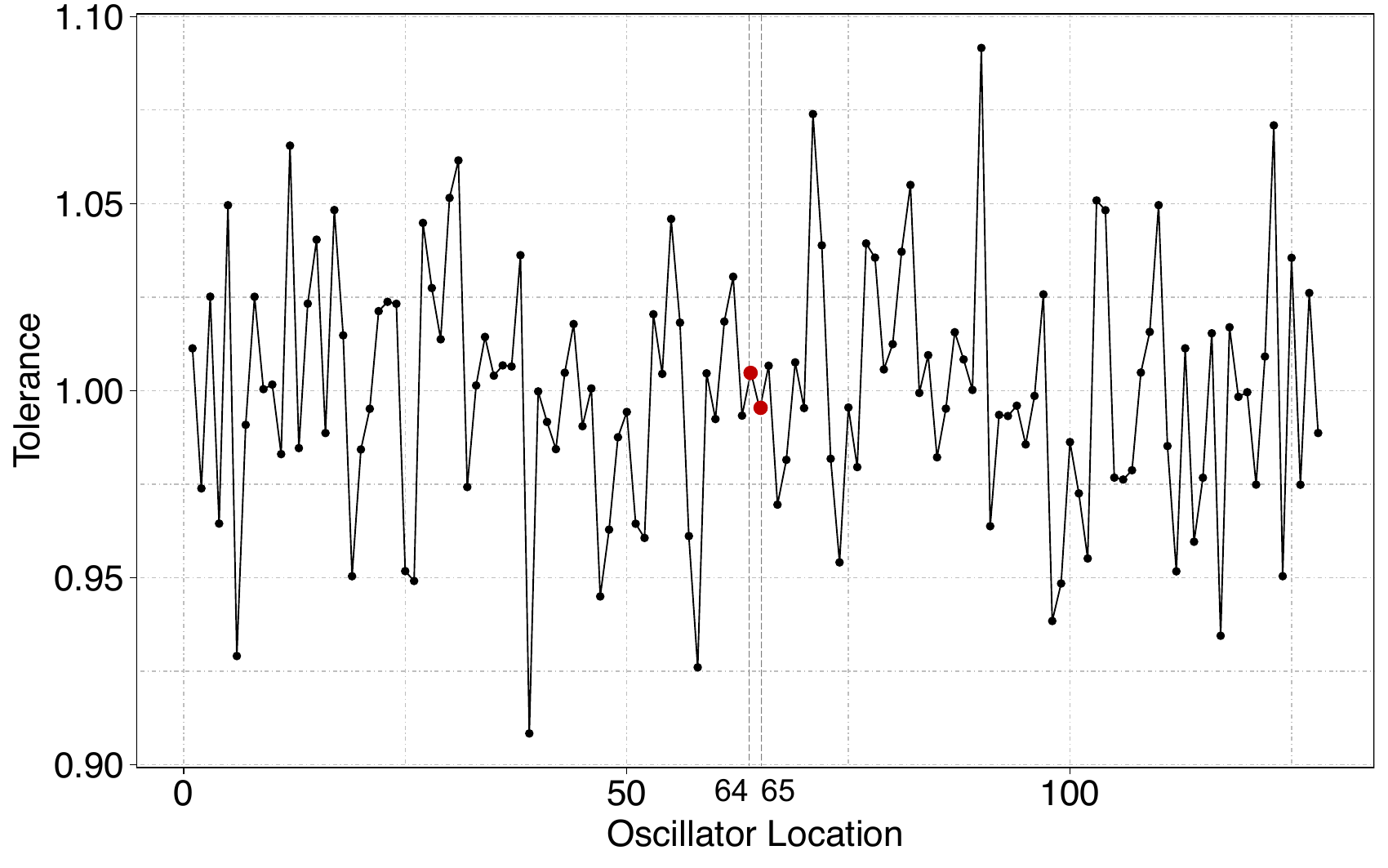}
	\caption{An example of a centrosymmetric distribution of 128 tolerance values.}
	\label{fig:centrosy}
\end{figure}

Energy transfer from the first mode to higher modes is symmetric with respect to the center location of a 1D FPUT-$\beta$ array (see Fig.~\ref{fig:energychange}). The configuration with oscillators ordered by ascending tolerance values is close to a centrosymmetric configuration. We hypothesize that centrosymmetry can lead to a recovery of recurrence. In a centrosymmetric $N$-oscillator array with even $N$, if the oscillator at location $N/2 - i$ has a parameter value that differs from the mean by $+\varpi$, then the associated parameter value of the oscillator at location $N/2 + i$ differs from the mean by $-\varpi$. In Fig.~\ref{fig:centrosy}, we show a centrosymmetric distribution for an array of $N = 128$ oscillators. This distribution has a mean of $1$ and centrosymmetry with respect to oscillators 64 and 65.

We conduct simulations to investigate the use of centrosymmetry {for} the recovery of recurrence in FPUT arrays with tolerance. We again use FPUT-$\beta$ arrays with $N = 128$ oscillators, a nonlinear coupling strength of $\beta = 8$, and a tolerance of $\pm 5\%$. With arbitrarily placed tolerance values from a Gaussian distribution, only 15 of our 100 simulations have notable energy recurrence. When we arrange the oscillators in ascending order of their tolerance values, 38 of 100 simulations have notable energy recurrence. When we arrange the oscillators centrosymmetrically, we observe energy recurrence in 76 of 100 simulations. The centrosymmetric distribution has a stronger effect on energy recurrence than ordering the oscillators by their tolerance values.

For a $\pm 10\%$ tolerance, only 3 of our 100 simulations with arbitrarily placed tolerance values have notable energy recurrence. However, rearranging the oscillators in ascending order of their tolerance values or using a centrosymmetric arrangement yields recurrence in 25 of 100 simulations in both cases.

This recovery is not as good as the recovery at a $\pm 5\%$ tolerance. Nevertheless, even with very large tolerance levels, both the centrosymmetric distribution and {ordering oscillators by ascending order of their tolerance values} can reduce the impact of tolerance on the energy recurrence {of} FPUT arrays.

Energy recurrence in FPUT arrays is not completely understood, and analyzing the effect of centrosymmetry on such recurrence provides a valuable perspective on FPUT systems. Centrosymmetry has also been an important consideration in studies of other systems. For example, Kosevich~\cite{Kosevich1993} investigated the effects of anharmonicity on transverse motion in a centrosymmetric 1D array of nonlinearly coupled elements. Moreover, in the arrays of molecules (so-called ``molecular networks'') that were investigated by Zech et al.~\cite{zech2014}, centrosymmetry
can mediate efficient, coherence-induced quantum transport. More recently, Chu and Yang~\cite{Chu2017} studied a two-dimensional ionic array of rocksalt type and observed using Monte Carlo simulations that flexoelectricity (i.e., the inducement of electric polarization by strain gradients) occurred only in noncentrosymmetric arrangements. Based both on our investigation and on these prior studies, it seems that centrosymmetry has interesting effects on nonlinear dynamics, although further investigation is necessary to fully understand it.


\section{Conclusions} \label{conc}

We studied 1D Fermi--Pasta--Ulam--Tsingou (FPUT) arrays with cubic nonlinear coupling between oscillators and tolerances in the oscillators. Increasing the tolerance, which increases the amount of heterogeneity of an array, breaks the energy recurrence of these FPUT-$\beta$ arrays. With tolerances, we also observed that energy can transfer to even modes in an FPUT-$\beta$ array; this cannot occur in the associated homogeneous FPUT system. Given the tolerance values of a set of oscillators, we rearranged them (1) by ordering them from smallest to largest and (2) centrosymmetrically, and we showed in both cases that such rearrangement can mitigate the loss of energy recurrence. We also observed that the tolerance values of individual oscillators can play a significant role in the breakdown of recurrence.


\section*{Acknowledgements}

ZL and BC thank {the} German Federal Ministry of Education and Research (Bundesministerium f\"ur Bildung und Forschung) and {the} European Commission for funding {from} the Distributed Artificial Intelligence Systems {(DAIS) project} (Grant agreement ID 101007273).

%


\end{document}